\documentclass[12pt,english,aps,manuscript]{article}
\usepackage[T1]{fontenc}
\usepackage[latin1]{luainputenc}
\usepackage{geometry}
\usepackage{amsmath}
\usepackage{amssymb}
\usepackage{setspace}
\usepackage[numbers]{natbib}
\makeatletter

\usepackage{geometry}

\makeatletter

\usepackage{color}

\makeatletter
\newcommand{\bee}{\begin{equation}}
\newcommand{\ee}{\end{equation}}
\newcommand{\beea}{\begin{eqnarray}}
\newcommand{\eea}{\end{eqnarray}}

\makeatother

\makeatother

\makeatother

\usepackage{babel}
\begin{document}
\begin{center}
\textbf{\Large{}Notes on LSZ, i$\boldsymbol{\epsilon}$ Prescriptions
and Perturbation Theory, in QFT and Cosmology}{\Large\par}
\par\end{center}

\begin{flushleft}
\vspace{0.3cm}
\par\end{flushleft}

\begin{center}
{\large{}S. P. de Alwis$^{\dagger}$ }{\large\par}
\par\end{center}

\begin{center}
Physics Department, University of Colorado, \\
 Boulder, CO 80309 USA 
\par\end{center}

\begin{center}
\vspace{0.3cm}
 
\par\end{center}

\begin{center}
\textbf{Abstract} 
\par\end{center}

\begin{center}
\vspace{0.3cm}
\par\end{center}

\smallskip{}
\vspace{0.3cm}
 We review the original argument of Lehmann et al (LSZ) that relates
the flat space S-matrix to the correlation function of field operators
and clarify some confusing issues. Next we discuss the origin of the
$i\epsilon$ prescriptions following Weinberg, but without assuming
that the vacuum is free at asymptotic times. Then we discuss the corresponding
argument in inflationary cosmology, emphasizing that unitarity violating
contour deformations are unnecessary.
\begin{center}
\vspace{0.3cm}
 
\par\end{center}

\vfill{}

$^{\dagger}$ dealwiss@colorado.edu

\eject

\section{Review of the LSZ theorem}

In textbook discussions it is often asserted that the LSZ theorem
\citep{Lehmann:1954rq} requires that asymptotically (i.e. in the
far past or the far future), the interaction between particles is
adiabatically turned off\footnote{For a recent discussion of this theorem, and an argument that strong
asymptotic limits in time are needed to precisely define particle
scattering, see \citep{Collins:2019ozc}. Our discussion is somewhat
complementary to that work.}. On the other hand, the particles in the asymptotic regions are supposed
to have physical masses, or in other words, the self interaction remains
non-zero. We point out below that the LSZ argument does not require
this rather unphysical (it violates time translation invariance),
assumption. 

As is the case in the LSZ paper we will for simplicity deal with a
scalar field theory with a single scalar field $\hat{\phi}\left(x\right)$
with physical mass $m$. LSZ make the following assumptions.
\begin{enumerate}
\item Poincare invariance.
\item Micro-causality, $\left[\hat{\phi}\left(x\right),\hat{\phi}\left(y\right)\right]=0,\,{\rm for}\,\left(x-y\right)\,{\rm space-like.}$
\item Asymptotic condition. The field operator $\hat{\phi}(x)\rightarrow\hat{\phi}_{\pm}(x)$
for $x^{0}\rightarrow\pm\infty$, where the asymptotic fields $\phi_{\pm}(x)$
propagate as free fields but with physical mass $m$. These are the
so-called out/in fields.
\end{enumerate}
While 1. and 2. are standard assumptions of a relativistic local QFT
in flat space, assumption 3. has been the cause of much confusion.

LSZ elaborate on 3. as follows. They introduce annihilation (and conjugate
creation) operators
\begin{equation}
\hat{\phi}^{f}(t)=i\int_{x^{0}=t}d^{3}x\left\{ \hat{\phi}(x)\frac{\partial f(x)}{\partial x^{0}}-f(x)\frac{\partial\hat{\phi}(x)}{\partial x^{0}}\right\} \equiv i\int_{x^{0}=t}d^{3}x\left\{ \hat{\phi}(x)\overleftrightarrow{\frac{\partial}{\partial x^{0}}}f(x)\right\} .\label{eq:phif}
\end{equation}
Here $f$ are normalized positive frequency solutions of the Klein-Gordon
equation (KG):
\begin{equation}
\left(\square-m^{2}\right)f=0,\,i\int_{x^{0}=t}d^{3}x\left\{ f(x)\overleftrightarrow{\frac{\partial}{\partial x^{0}}}f^{*}(x)\right\} =1\label{eq:feqns}
\end{equation}
Then for any normalizable states $\Phi,\Psi$ the asymptotic conditions
are taken to be
\begin{equation}
\lim_{t\rightarrow\pm\infty}<\Phi|\hat{\phi}^{f}(t)|\Psi>=<\Phi|\hat{\phi}_{\pm}^{f}(t)|\Psi>,\label{eq:asymp}
\end{equation}
and it is asserted that the RHS is independent of t and so is constant.
It is important that these conditions are formulated as limits of
matrix elements between normalizable states rather than as operator
limits (as naively stated in assumption 3. above). The latter, i.e.
the strong limit as opposed to the weak limit in eqn. \eqref{eq:asymp},
would imply that the theory is free.

Note that for $f$ a positive frequency solution of the KG equation,
$\hat{\phi}_{\pm}^{f}$ is an annihilation operator whose asymptotic
limits give free field like annihilation operators. So LSZ assert
that there are states $\Omega_{\pm}$ (in and out vacuum states) such
that
\begin{equation}
\hat{\phi}_{\pm}^{f}|\Omega_{\pm}>=0,\,{\rm with\,\Omega_{+}=\Omega_{-}=\Omega,}\label{eq:Omegapm}
\end{equation}
In other words there is no distinction between these in and out states,
and furthermore they are in fact equal to the vacuum state of the
interacting theory\footnote{So $\Omega_{\pm}$ should not be identified with the vacuum of the
free theory.} with 4-momentum operator $P_{\mu}$, and is an eigenstate with zero
energy-momentum,
\[
P_{\mu}|\Omega>=0.
\]

We will show using the Riemann-Lebegues (RL) theorem that equation
\eqref{eq:asymp} is actually true - in other words it is not an assumption.

Since $f$ is a positive frequency solution of the K-G eqn. we can
write in Fourier space,
\begin{equation}
f(x)=\int\widetilde{dp}e^{-iE_{p}t+i{\bf p}.{\bf x}}\tilde{f}({\bf p}),\,E_{p}=\sqrt{{\bf p}^{2}+m^{2}},\,\widetilde{dp}=:\frac{d^{3}p}{\left(2\pi\right)^{3}2E_{p}}.\label{eq:fFT}
\end{equation}
We note in passing that since $f(x)$ is normalized (in the KG norm)
see \eqref{eq:feqns}, we have $(f,f)_{{\rm KG}}=\int\widetilde{dp}|\tilde{f}({\bf p})|^{2}=1$.
The Fourier transform of the quantum field is
\begin{equation}
\hat{\phi}(x)=\int d^{4}q\left\{ e^{-iqx}\tilde{\phi}(q)\right\} ,\,\,\tilde{\phi}^{*}(q)=\tilde{\phi}(-q).\label{eq:phiFT}
\end{equation}
Now let us separate this in Fourier space into a part which satisfies
the homogeneous field equation (i.e the KG eqn.) and a piece which
is interaction dependent, i.e. would be zero if the theory is free.
So we write\footnote{$\phi_{I}(x)$ is taken to be the advanced or retarded solution depending
on the asymptotic solution we are interested in. Here we suppress
these distinctions.}
\begin{equation}
\tilde{\phi}(q)=\tilde{\phi}_{0}(q)\delta^{+}\left(q^{2}+m^{2}\right)+\tilde{\phi}_{I}\left(q\right).\label{eq:phitildeq}
\end{equation}
Substituting this into \eqref{eq:phiFT} we have
\begin{equation}
\phi(x)=\int\frac{d^{3}q}{2E_{q}}\left(e^{iE_{q}t-i{\bf q}\cdot{\bf x}}\tilde{\phi}_{0}(q)+{\rm c.c}\right)+\int d^{4}q\left\{ e^{-iqx}\tilde{\phi}_{I}(q)\right\} .\label{eq:phisplit}
\end{equation}
Computing the two terms of \eqref{eq:phif} we have, after doing the
spatial integration,
\begin{align*}
i\int d^{3}x\partial_{t}f(x)\phi(x) & =\frac{1}{2}\int d^{3}q\tilde{f}\left({\bf q}\right)\tilde{\phi}_{0}({\bf q})+\frac{1}{2}\int d^{4}qe^{-i\left(E_{q}-q^{0}\right)t}\tilde{f}({\bf q})\tilde{\phi}_{I}({\bf q},q^{0}),\\
-i\int d^{3}xf(x)\partial_{t}\phi(x) & =\frac{1}{2}\int d^{3}q\tilde{f}\left({\bf q}\right)\tilde{\phi}_{0}({\bf q})+\frac{1}{2}\int d^{4}qe^{-i\left(E_{q}-q^{0}\right)t}\frac{q^{0}}{E_{q}}\tilde{f}({\bf q})\tilde{\phi}_{I}({\bf q},q^{0}).
\end{align*}
So when taking the limit $t\rightarrow\pm\infty$ of the above sandwiched
between any two normalizable states $\Phi,\Psi$ we see that the second
terms in the two equations go to zero (faster than any power since
the factors multiplying the exponential are assumed to be $C_{\infty}$
functions as usual), by the Riemann-Lebesgue theorem\footnote{It should be emphasized that although the integration in the second
term of these equations has a point where $q_{0}=E_{q}$ at which
the $t$ dependence vanishes, this is a set of measure zero and will
not contribute to the integral. The first term separates because it
comes from the delta function term in \eqref{eq:phitildeq} .}. Hence we are just left with the sum of the first terms. i.e. (restoring
the distinction between the two asymptotic solutions in \eqref{eq:phitildeq}),
\begin{equation}
\lim_{t\rightarrow\pm\infty}<\Phi|\hat{\phi}_{f}(t)|\Psi>=<\Phi|\int d^{3}q\tilde{f}\left({\bf q}\right)\tilde{\phi}_{0\pm}({\bf q})|\Psi>.\label{eq:asymp1}
\end{equation}
Comparing with \eqref{eq:asymp} we may identify the in/out creation
operators as
\begin{equation}
\hat{\phi}_{\pm}^{f}(t)=\int d^{3}q\tilde{f}\left({\bf q}\right)\tilde{\phi}_{0\pm}({\bf q}),\label{eq:pbifpm}
\end{equation}
which are clearly time independent.

This justifies the LSZ assumption 3. The rest of the proof goes through
as in the original paper. In the text book by Sredinicki \citep{Srednicki:2007qs},
the RL theorem is also used, but for somewhat different reasons, namely
to argue that the asymptotic operators create only one-particle states
from the vacuum. In the argument in these notes this is guaranteed
by the explicit construction above.

\section{Perturbation Theory and the $i\epsilon$ Prescription}

In some textbook discussions of this it is suggested that the transition
from the Feynman kernel (which is the natural object for which the
functional integral is derived) to the correlation function of field
operators in the vacuum, is achieved by deforming the time direction
into the complex plane. This would seem to violate unitarity. We show
how to avoid this by slightly generalizing an argument of Weinberg
\citep{Weinberg:1995mt}\footnote{See section 9.2.}.

For simplicity let us again work with a single (real) scalar field
$\phi$, with conjugate momentum $\pi.$The functional integral for
a theory with Hamiltonian density ${\cal H}$ for the amplitude for
a transition from a field eigenstate $|\phi_{-}>$ in the far past
$t=-T\rightarrow-\infty$ to a field eigenstate $|\phi_{+}>$ in the
far future $t=+T\rightarrow\infty$, in the presence also of an external
source term $j(x)\phi(x)$ is
\begin{align}
\lim_{T\rightarrow\infty} & <\phi_{+}|Te^{i\int_{-T}^{+T}dt\int d^{3}x\left\{ H(\phi,\pi)+j\phi\right\} }|\phi_{-}>=\lim_{T\rightarrow\infty}\int_{\phi(-T)=\phi_{-}}^{\phi(T)=\phi_{+}}[d\phi][d\pi]e^{i\int_{-T}^{T}dt\int d^{3}x\left(\pi\dot{\phi}-{\cal H+}j\phi\right)}\nonumber \\
 & =\lim_{T\rightarrow\infty}\int_{\phi(-T)=\phi_{-}}^{\phi(T)=\phi_{+}}[d\phi]e^{i\int_{-T}^{T}dt\int d^{3}x\left({\cal L+}j\phi\right)}.\label{eq:PI}
\end{align}
Here the last line obtains if the Hamiltonian is no more than quadratic
in the conjugate momentum $\pi$. Functional derivatives of this object
will give time ordered correlation functions of the field operators
between eigenstates of the field operator, namely $|\phi_{-}>,|\phi_{+}>$.
However what we need to compute, in order to get the S-matrix using
the LSZ formula, is the vacuum to vacuum correlation function. For
this we need to multiply the above functional integral by the vacuum
wave functions $\Omega(\phi_{-})=<\phi_{-}|\Omega>,\,\Omega^{*}\left(\phi_{+}\right)=<\Omega|\phi_{+}>$,
and integrate over all values of $\phi_{-},\phi_{+}$. Then we get
for the generating functional for vacuum to vacuum correlation functions
the following expression,
\begin{equation}
e^{iW\left\{ j\right\} }=\int[d\phi]\Omega^{*}\left(\phi_{+}\right)\Omega(\phi_{-})e^{i\int dt\int d^{3}x\left({\cal L+}j\phi\right)}\label{eq:Wj}
\end{equation}
where now the functional integral includes all values of $\phi,$i.e.
including $\phi\left(\pm T\right)$, and we've taken the limit $T\rightarrow\infty$.
Let us now split the Lagrangian as well as the wave functions into
a free and interacting part. So we write,
\begin{align}
L=L_{0}+L_{I},\,\Omega\left[\phi_{\pm}\right] & =\exp\left\{ \omega_{0}^{\pm}+\omega_{I}^{\pm}\right\} ,\label{eq:LOmega}\\
L_{0}=-\int d^{3}x\frac{1}{2}\left(\partial\phi\partial\phi+m^{2}\phi^{2}\right) & ,\,\,\omega_{0}^{\pm}=-\int d^{3}x\int d^{3}y\frac{1}{2}\phi\left({\bf x},\pm\infty\right)k\left({\bf x},{\bf y}\right)\phi\left({\bf y},\pm\infty\right),\label{eq:L0omega0}\\
k\left({\bf x},{\bf y}\right) & =\left(2\pi\right)^{-3}\int d^{3}pe^{i\left({\bf x}-{\bf y}\right)\cdot{\bf p}}\sqrt{{\bf p}^{2}+m^{2}}.\label{eq:k}
\end{align}
Here $\omega_{I}$ contains cubic and higher powers of the asymptotic
fields and $\exp\left\{ \omega_{0}^{\pm}\right\} $ is the free wave
function, essentially a product of harmonic oscillator wave functions.
Now we observe that 
\begin{equation}
\Omega^{*}\left(\phi_{+}\right)\Omega(\phi_{-})=e^{\omega(+\infty)+\omega\left(-\infty\right)}=\lim_{\epsilon\rightarrow0}\exp\left[\epsilon\int_{-\infty}^{\infty}dte^{-\epsilon|t|}\left(\omega_{0}(t)+\omega_{I}\left(t\right)\right)\right]\label{eq:Omegaformula}
\end{equation}
As shown in \citep{Weinberg:1995mt} the free part of the wave function
just gives the $i\epsilon$ prescription for the propagator. Let us
rewrite the free part of the Lagrangian in \eqref{eq:Wj} by including
the contribution from $\omega_{0}$ using \eqref{eq:Omegaformula}
which is explicitly $\epsilon\int dt\int d^{3}x\int d^{3}y\phi({\bf x},t)k\left({\bf x},{\bf y}\right)\phi({\bf y},t)$.
Thus we replace $S_{0}$ by (renaming $\left({\bf y},t'\right)$ as
$x'$) 
\begin{align}
-\int d^{4}xd^{4}x'\phi(x)\left\{ \left(\partial_{\mu}\partial'^{\mu}+m^{2}\right)\delta\left(x-x'\right)-i\epsilon k\left({\bf x},{\bf x'}\right)\delta\left(t-t'\right)\right\} \phi\left(x'\right)\nonumber \\
:=-\int d^{4}xd^{4}x'\phi(x)K_{\epsilon}(x,x')\phi\left(x'\right) & .\label{eq:S_0}
\end{align}
Rewriting the interaction part of the functional integral in \eqref{eq:Wj}
by replacing the fields by $i^{-1}\delta/\delta j$ taken outside
the functional integral and performing the Gaussian integral over
$\phi$ (a standard manipulation) we get,
\begin{equation}
e^{iW\left\{ j\right\} }={\rm lim_{\epsilon\rightarrow0}}e^{i\int d^{4}x\left\{ {\cal L}_{I}\left(\frac{\delta}{\delta j(x)}\right)-i\epsilon\omega_{I}\left(\frac{\delta}{\delta j(x)}\right)\right\} }\exp\left\{ -i\frac{1}{2}\int d^{4}xd^{4}x'j(x)K_{\epsilon}^{-1}\left(x,x'\right)j(x')\right\} .\label{eq:jk-1K}
\end{equation}
Here written as a Fourier integral
\begin{equation}
K_{\epsilon}^{-1}\left(x,x'\right)=\left(2\pi\right)^{-4}\int d^{4}pe^{i\left(x-x'\right)p}(p^{2}+m^{2}-i\epsilon E_{p})^{-1},\label{eq:K^-1}
\end{equation}
($E_{p}=\sqrt{{\bf p}^{2}+m^{2}}$). This is simply the Feynman propagator.
Now note also that the interaction part of the wave function (i.e.
the $i\epsilon\omega_{I}$ term outside the second exponential) has
no effect since it disappears in the $\epsilon\rightarrow0$ limit.
So the only contribution of the wave functions (even including the
interaction) is to give the standard formula for deriving perturbation
theory with the $i\epsilon$ prescription.

All that we've done here is to show that one should actually use the
full interacting vacuum wave function and one need not argue as Weinberg
and others do that the interaction turns off at asymptotic times so
that the interacting vacuum can be replaced by the free vacuum. The
final formula is the same but the point is that now we've established
the perturbation theory for computing correlation functions in the
\textit{interacting} vacuum as is required by LSZ.

\section{Cosmological Perturbation Theory and the $i\epsilon$ Prescription}

In cosmology one needs the equal time correlation function of field
operators at some late time $t$ expressed in term of the same evaluated
in the far past. In the literature the state in the far past is argued
to be the free Bunch Davies vacuum. The justification for this (as
well as getting the $i\epsilon$ prescription for perturbation theory)
is usually made by making a contour rotation of the time integral
into the imaginary direction (see for example \citep{Baumann:2014nda}
appendix C.3 ). Below we will show that this (unitarity violating)
construction is unnecessary, by generalizing the argument in the previous
section\footnote{An alternative procedure for avoiding unitarity violation had been
proposed in \citep{Kaya:2018jdo}. This amounts to redefining the
interaction Hamiltonian by multiplying by powers of factor $\exp(\epsilon t)$
. Again the asymptotic limit is taken before $\epsilon$ is taken
to zero so this is tantamount to taking the interaction to zero in
the asymptotic past. This violates time translations in flat space
and in cosmology where$\frac{dH}{dt}=\frac{\partial H}{\partial t}\propto\dot{a}$
there is an additional term proportional to $\epsilon$.}.

It is convenient to derive the relevant formulae for the general case
that obtains in statistical field theory/non-equilibrium QFT. Here
one is interested in the expectation value of correlation functions
of Heisenberg picture fields in a state given by a density matrix
$\rho_{0}$. One may define then the generator of connected correlation
functions $W\left[J.\rho_{0}\right]$ as 
\begin{equation}
\exp\left(iW\left[J.\rho_{0}\right]\right)={\rm tr}\left\{ \rho_{0}P_{C}\exp\left(i\int_{C}dt'\int d^{3}xJ\left({\bf x},t'\right)\Phi\left({\bf x},t'\right)\right)\right\} .\label{eq:WJrho}
\end{equation}
Here $J$ is a classical source and $\Phi$ is the field operator
(for simplicity of presentation we'll just focus on a single scalar
field). The contour $C$ starts at large negative time $-T$ proceeds
to $t$ ($C_{+}$) and then reverses direction and goes back to $-T$,
($C_{-})$. $P_{C}$ is an instruction to path order along the contour
$C$. Note that it is this path ordering that makes the RHS non-trivial.
So along $C_{+}$ we have time ordering while along $C_{-}$ we have
anti-time ordering. It is useful to rewrite the exponent on the RHS
of \eqref{eq:WJrho} as
\[
i\int_{C}dt'\int d^{3}xJ\left({\bf x},t'\right)\hat{\Phi}\left({\bf x},t'\right)=i\int_{-T}^{t}dt'\int d^{3}xJ^{+}\left({\bf x},t'\right)\hat{\Phi}\left({\bf x},t'\right)-i\int_{-T}^{t}dt'\int d^{3}xJ^{-}\left({\bf x},t'\right)\hat{\Phi}\left({\bf x},t'\right)
\]
So for the connected two point functions we have four possibilities,
\begin{align}
\frac{\delta^{2}iW}{i\delta J^{+}\left(x\right)i\delta J^{+}\left(y\right)} & =<T\left\{ \hat{\Phi}\left(x\right)\hat{\Phi}\left(y\right)\right\} >_{0}^{x^{0},y^{0}\in C^{+}},\,\frac{\delta^{2}iW}{i\delta J^{-}\left(x\right)i\delta J^{-}\left(y\right)}=<\overline{T}\left\{ \hat{\Phi}\left(x\right)\hat{\Phi}\left(y\right)\right\} >_{0}^{x^{0},y^{0}\in C^{-}},\label{eq:W++--}\\
\frac{\delta^{2}iW}{i\delta J^{+}\left(x\right)i\delta J^{-}\left(y\right)} & =<\hat{\Phi}\left(y\right)\hat{\Phi}\left(x\right)>_{0}^{x^{0}\in C^{+},y^{0}\in C^{-}},\,\frac{\delta^{2}iW}{i\delta J^{-}\left(x\right)i\delta J^{+}\left(y\right)}=<\hat{\Phi}\left(x\right)\hat{\Phi}\left(y\right)>_{0}^{x^{0}\in C^{-},y^{0}\in C^{+}}.\label{eq:W+--+}
\end{align}
Here $<*>_{0}:={\rm tr}(*\rho_{0})$. Calling the four two point functions
$G^{++},G^{--},G^{+-},G^{-+}$ in an obvious notation, we have the
relation,

\[
G^{++}+G^{--}=G^{+-}+G^{-+}.
\]

The most convenient way of evaluating \eqref{eq:WJrho} is by going
to its functional integral representation. To do so evaluate the trace
in the eigenbases of the field operators in the upper($\Phi^{+}$)/lower($\Phi^{-})$
branch of $C$ at time $-T$, (with $T\rightarrow\infty$)
\begin{equation}
\Phi^{\pm}(-T,{\bf x})|\phi^{\pm}>=\phi^{\pm}({\bf x})|\phi^{\pm}>\label{eq:phibasis}
\end{equation}
 Using the completeness relation in the $\phi^{-}$ basis \eqref{eq:WJrho}
is,

\[
\exp\left(iW\left[J.\rho_{0}\right]\right)=\int\left[d\phi^{+}\right]\left[d\phi^{-}\right]<\phi^{+}|\rho_{0}|\phi^{-}><\phi^{-}|P_{C}\exp\left(i\int_{C}dt'\int d^{3}xJ\left({\bf x},t'\right)\hat{\Phi}\left({\bf x},t'\right)\right)|\phi^{+}>
\]
Here $[d\phi^{\pm}]:=\prod_{{\bf x}}d\phi^{\pm}({\bf x})$. The second
factor in the integrand above is just the transition amplitude to
go from $\phi^{+}$ at $-T$ to $\phi^{-}$ at $-T$ along the contour
$C$ in the presence of the external source $J$. This can now be
replaced by a path integral using the standard construction except
that the time slicing involves two branches: one going from $-T$
to $t$ (the upper branch) and a second going backwards from $t$
to $-T$ (the lower branch). Thus we get the functional integral representation
(if the dynamics is governed by a classical action $S\left(\phi\right)$)
\begin{equation}
\exp\left(iW\left[J.\rho_{0}\right]\right)=\int[d\Phi]_{C}<\phi^{+}|\rho_{0}|\phi^{-}>e^{i\left\{ S_{C}[\Phi]+\int_{C}dt'\int d^{3}xJ(x)\Phi(x)\right\} }\label{eq:WPI}
\end{equation}
Here the functional integral measure extends over all configurations
from $\Phi({\bf x},-T)$ to $\Phi({\bf x},t)$ along $C_{+}$ and
then back along $C_{-}$ to $\Phi({\bf x},-T)$. Now this expression
apart from the end points of the integrations is essentially the same
as that \eqref{eq:Wj} used to compute flat space correlation functions
which are needed for the S-matrix. Thus one would expect to get essentially
the same result, namely an equation similar to \eqref{eq:jk-1K}.
Let us spell out the differences. Firstly the vacuum density matrix
is
\[
<\phi^{+}|\rho_{0}|\phi^{-}>=\Omega[\phi^{+}]\Omega^{*}\left[\phi^{-}\right]
\]
 where $\phi^{\pm}$ are the end points at time $-T$ on the upper/lower
branch of the integration. So for the first factor we have by similar
reasoning as in the sequence of steps from \eqref{eq:LOmega} to\eqref{eq:Omegaformula}
\begin{equation}
\Omega[\phi^{+}]=e^{\omega[\Phi^{+},-\infty)}=\lim_{\epsilon\rightarrow0}\exp\left[\epsilon\int_{-\infty}^{t}dt'e^{-\epsilon t'}\left(\omega_{0}(t',\Phi^{+}(t')]+\omega_{I}(t',\Phi^{+}(t')]\right)\right]\label{eq:Omega+}
\end{equation}
with a similar expression for $\Omega^{*}\left[\phi^{-}\right]$ with
$\Phi^{+}\rightarrow\Phi^{-}$. Note the upper limit of the $t'$
integration is now $t$, not $+\infty$. Again $\omega_{0}$ is no
more than quadratic in the fields while $\omega_{I}$ contains cubic
and higher orders. We use these expressions for the wave functions
in \eqref{eq:WPI} and split the action into a free part (i.e. quadratic
in the fields) and an interaction (cubic and higher in the fields)
as in the previous section. For the free part we have (note that both
the free part and the interaction may depend explicitly on time as
in cosmology we have a time-dependent background), for the upper branch,
\begin{align*}
\int_{-\infty}^{t}dt'\int_{-\infty}^{t}dt''\int d^{3}xd^{3}x'\Phi^{+}(x)\left\{ K({\bf x},{\bf x'};t',t'')+i\epsilon k\left({\bf x},{\bf x'};t,t'\right)\delta\left(t'-t''\right)\right\} \Phi^{+}\left(x'\right)\\
:=\int_{-\infty}^{t}dt'\int_{-\infty}^{t}dt''\int d^{3}xd^{3}x'\Phi^{+}(x)K_{\epsilon}(x,x')\Phi^{+}\left(x'\right),
\end{align*}
where $K$ is the differential operator (explicitly time-dependent
in cosmology) that replaces the Klein-Gordon operator in equation
(\ref{eq:S_0}) and $k$ is the matrix which replaces \eqref{eq:k}
in the quadratic form for $\omega_{0}$. We have a similar expression
for the lower branch. Hence as before by using the replacement $\Phi^{\pm}\rightarrow\delta/\delta J^{\pm}$
we have
\begin{equation}
e^{iW\left\{ J\right\} }=\lim e_{\epsilon\rightarrow0}^{i\int_{C}d^{4}x\left\{ {\cal L}_{I}\left(\frac{\delta}{\delta J(x)}\right)-i\epsilon\omega_{I}\left(\frac{\delta}{\delta J(x)}\right)\right\} }\int[d\Phi]_{C}e^{i\int_{C}d^{4}x\int_{C}d^{4}x'\Phi(x)K_{\epsilon}(x,x')\Phi\left(x'\right)+\int_{C}d^{4}xJ(x)\Phi\left(x\right)}\label{eq:WJ2}
\end{equation}
The $\epsilon\rightarrow0$ suppresses the $\omega_{I}$term in the
exponential outside the functional integral as before. The latter
is just the free theory one for the closed time path. However this
integral is not a product over two Gaussian integrals one over $C_{+}$
and other over $C_{-}$. The two integrations are identified at $t'=t$.
Following the discussion in Calzetta and Hu \citep{Calzetta:2008iqa}
we will evaluate it by noticing that the free theory generating function
of connected correlation functions is, (for simplicity we'll impose
$<\Phi>_{{\rm free}}=0,$ which can be ensured by adding a linear
term to the action),
\begin{equation}
W_{0}[J]=-\frac{1}{2}\int_{-\infty}^{t}d^{4}xJ_{A}\left(x\right)G_{0}^{AB}J_{B}\left(x\right),\,A,B=\pm.\label{eq:W0}
\end{equation}
Here 
\begin{align}
G_{0}^{++}(x,y) & =Z_{0}^{-1}\int[d\Phi]_{C}e^{i\int_{C}d^{4}x'\int_{C}d^{4}x''\Phi^{+}(x')K_{\epsilon}(x',x'')\Phi^{+}\left(x''\right)}\Phi^{+}\left(x\right)\Phi^{+}\left(y\right)\label{eq:G++}\\
G_{0}^{--}(x,y) & =Z_{0}^{-1}\int[d\Phi]_{C}e^{i\int_{C}d^{4}x'\int_{C}d^{4}x''\Phi^{+}(x')K_{\epsilon}(x',x'')\Phi^{+}\left(x''\right)}\Phi^{-}\left(x\right)\Phi^{-}\left(y\right)\label{eq:G--}\\
G_{0}^{+-}(x,y) & =Z_{0}^{-1}\int[d\Phi]_{C}e^{i\int_{C}d^{4}x'\int_{C}d^{4}x''\Phi^{+}(x')K_{\epsilon}(x',x'')\Phi^{+}\left(x''\right)}\Phi^{+}\left(x\right)\Phi^{-}\left(y\right)\label{eq:G+-}\\
G_{0}^{-+}(x,y) & =Z_{0}^{-1}\int[d\Phi]_{C}e^{i\int_{C}d^{4}x'\int_{C}d^{4}x''\Phi^{+}(x')K_{\epsilon}(x',x'')\Phi^{+}\left(x''\right)}\Phi^{-}\left(x\right)\Phi^{+}\left(y\right)\label{eq:G-+}
\end{align}
Note that these are the free field correlation functions i.e. 
\begin{align*}
G_{0}^{++}(x,y) & =<T\left\{ \hat{\Phi}\left(x\right)\hat{\Phi}\left(y\right)\right\} >_{{\rm free}}^{x^{0},y^{0}\in C^{+}},\,G_{0}^{--}(x,y)=<\overline{T}\left\{ \hat{\Phi}\left(x\right)\hat{\Phi}\left(y\right)\right\} >_{{\rm free}}^{x^{0},y^{0}\in C^{-}}\\
G_{0}^{+-}(x,y) & =<\left\{ \hat{\Phi}\left(y\right)\hat{\Phi}\left(x\right)\right\} >_{{\rm free}}^{x^{0}\in C^{+},y^{0}\in C^{-}},\,G_{0}^{-+}(x,y)=<\left\{ \hat{\Phi}\left(x\right)\hat{\Phi}\left(y\right)\right\} >_{{\rm free}}^{x^{0}\in C^{-},y^{0}\in C^{+}}
\end{align*}
If we are dealing with a flat space relativistic theory then the ones
on the first line are just the Feynman and Dyson propagators (i.e.
the complex conjugate of the Feynman propagator \eqref{eq:K^-1}).
In cosmology the relevant free two-point functions would be the corresponding
ones in (quasi)deSitter space with the different boundary conditions.
However the $i\epsilon$ prescription is the same as in flat space
and our discussion establishes cosmological perturbation theory for
equal time correlation functions evaluated in the interacting vacuum.

\section{Conclusions}

In this paper we have discussed calculations of both flat space S-matrices
and equal time correlators in cosmology using perturbation theory,.
What we have shown is that there is no need to make the assumptions
that are usually made, either that at asymptotic times the interaction
is turned off, or that it is effectively turned off because of a contour
deformation of the time integral in an imaginary direction. The first
assumption would violate time translation invariance (in addition
to what is required by the cosmological background), while the second
violates unitarity. In our functional integral derivation of perturbation
theory we have kept the in-out (for the S-matrix) or in-in (for cosmology)
of the asymptotic states, to be the wave function solutions of the
full interacting theory. Nevertheless the non-linear piece (in the
log of the wave function), which can be factored out using a standard
field theory technique, has no effect on the calculation. This justifies
the usual calculations without any symmetry/unitarity violating assumptions.

\section{Acknowledgements}

I wish to thank Sebastian Cespedes and Fernando Quevedo for useful
discussions and comments. I also wish to thank Tom DeGrand for a careful
reading of the manuscript.

\bibliographystyle{apsrev}
\bibliography{myrefs}

\end{document}